\documentclass[aps,prl,twocolumn,showpacs,letter,superscriptaddress]{revtex4}
\usepackage{graphicx}

\begin{document}

\title{Attosecond electron spectroscopy using a novel interferometric pump-probe technique}

\author{J.~Mauritsson}
\affiliation{Department of Physics, Lund Institute of Technology, P.
O. Box 118, SE-221 00 Lund, Sweden}

\author{T.~Remetter}
\affiliation{Department of Physics, Lund Institute of Technology, P.
O. Box 118, SE-221 00 Lund, Sweden}

\author{M.~Swoboda}
\affiliation{Department of Physics, Lund Institute of Technology, P.
O. Box 118, SE-221 00 Lund, Sweden}

\author{K.~Kl\"under}
\affiliation{Department of Physics, Lund Institute of Technology, P.
O. Box 118, SE-221 00 Lund, Sweden}

\author{A.~L'Huillier}
\affiliation{Department of Physics, Lund Institute of Technology, P.
O. Box 118, SE-221 00 Lund, Sweden}

\author{K.~J.~Schafer}
\affiliation{Department of Physics and Astronomy, Louisiana State
University, Baton Rouge, Louisiana 70803-4001, USA}

\author{O.~Ghafur}
\affiliation{FOM-Institute AMOLF, Science park 113, 1098 XG Amsterdam, The Netherlands}

\author{F.~Kelkensberg}
\affiliation{FOM-Institute AMOLF, Science park 113, 1098 XG Amsterdam, The Netherlands}

\author{W.~Siu}
\affiliation{FOM-Institute AMOLF, Science park 113, 1098 XG Amsterdam, The Netherlands}

\author{P.~Johnsson}
\affiliation{FOM-Institute AMOLF, Science park 113, 1098 XG Amsterdam, The Netherlands}

\author{M.~J.~J.~Vrakking}
\affiliation{FOM-Institute AMOLF, Science park 113, 1098 XG Amsterdam, The Netherlands}
\affiliation{Max-Born-Institut f\"ur Nichtlineare Optik und Kurzzeitspektroskopie (MBI), Max-Born-Straße 2 A, 12489 Berlin, Germany}

\author{I.~Znakovskaya}
\affiliation{4 Max-Planck-Institut f\"ur Quantenoptik, Hans-Kopfermann-Strasse 1, D-85748 Garching, Germany}

\author{T.~Uphues}
\affiliation{4 Max-Planck-Institut f\"ur Quantenoptik, Hans-Kopfermann-Strasse 1, D-85748 Garching, Germany}

\author{S.~Zherebtsov}
\affiliation{4 Max-Planck-Institut f\"ur Quantenoptik, Hans-Kopfermann-Strasse 1, D-85748 Garching, Germany}

\author{M.~F.~Kling}
\affiliation{4 Max-Planck-Institut f\"ur Quantenoptik, Hans-Kopfermann-Strasse 1, D-85748 Garching, Germany}

\author{F.~L\'epine}
\affiliation{Université Lyon 1; CNRS; LASIM, UMR 5579, 43 bvd. du 11 novembre 1918, F-69622 Villeurbanne, France}

\author{E.~Benedetti}
\affiliation{Politecnico di Milano, Department of Physics Istituto di Fotonica e Nanotecnologie, CNR-IFN Piazza L. da Vinci 32, 20133 Milano, Italy}

\author{F.~Ferrari}
\affiliation{Politecnico di Milano, Department of Physics Istituto di Fotonica e Nanotecnologie, CNR-IFN Piazza L. da Vinci 32, 20133 Milano, Italy}

\author{G.~Sansone}
\affiliation{Politecnico di Milano, Department of Physics Istituto di Fotonica e Nanotecnologie, CNR-IFN Piazza L. da Vinci 32, 20133 Milano, Italy}

\author{M.~Nisoli}
\affiliation{Politecnico di Milano, Department of Physics Istituto di Fotonica e Nanotecnologie, CNR-IFN Piazza L. da Vinci 32, 20133 Milano, Italy}

\begin{abstract}
We present an interferometric pump-probe technique for the characterization of attosecond electron wave packets (WPs) that uses a free WP as a reference to measure a bound WP. We demonstrate our method by exciting helium atoms using an attosecond pulse with a bandwidth centered near the ionization threshold, thus creating both a bound and a free WP simultaneously. After a variable delay, the bound WP is ionized by a few-cycle infrared laser precisely synchronized to the original attosecond pulse. By measuring the delay-dependent photoelectron spectrum we obtain an interferogram that contains both quantum beats as well as multi-path interference. Analysis of the interferogram allows us to determine the bound WP components with a spectral resolution much better than the inverse of the attosecond pulse duration.
\end{abstract}

\pacs{32.80.Rm, 32.80.Qk, 42.65.Ky}

\maketitle

Attosecond science~\cite{PaulScience2001,HentschelNature2001} promises to achieve temporal resolution comparable to the duration ($\tau$) of the light pulses, \textit{e.g.}, of the order of 100\,as~\cite{SansoneScience2006,GustafssonOL2007} or even below~\cite{GoulielmakisScience2008}. An important issue, however, is whether this is possible only to the detriment of spectral resolution, thus considerably limiting the scientific interest of such light sources. When attosecond pulses interact with atoms or molecules, they create broad electron wave packets (WPs), partly in the continuum, but often including also a number of bound states excited by direct absorption~\cite{JohnssonPRL2007} and/or by shake-up processes~\cite{UiberackerNature2007}. A spectral resolution given by the Fourier limit, \textit{i.e.}, of the order of $1/\tau$, prevents any detailed analysis of such complex WPs.

Because of the low intensity of currently available attosecond pulses, most of the techniques used to characterize attosecond electron WPs have used as a probe an infrared (IR) pulse which is synchronized to the extreme ultraviolet (XUV) attosecond pulse. The required XUV-IR synchronization is inherent to the attosecond pulse generation process and several methodologies, such as chronoscopy~\cite{UiberackerNature2007}, streaking~\cite{KienbergerScience2002,ItataniPRL2002}, stroboscopy~\cite{MauritssonPRL2008} and interferometry~\cite{RemetterNP2006} have been demonstrated. All of these techniques require that the attosecond and IR pulses overlap temporally. The IR field is therefore not only probing the electron WP after it has been created but it also perturbs the formation. In addition, the spectral resolution of these techniques is Fourier limited.

\begin{figure}[t]\centering
\includegraphics[width=0.65\linewidth]{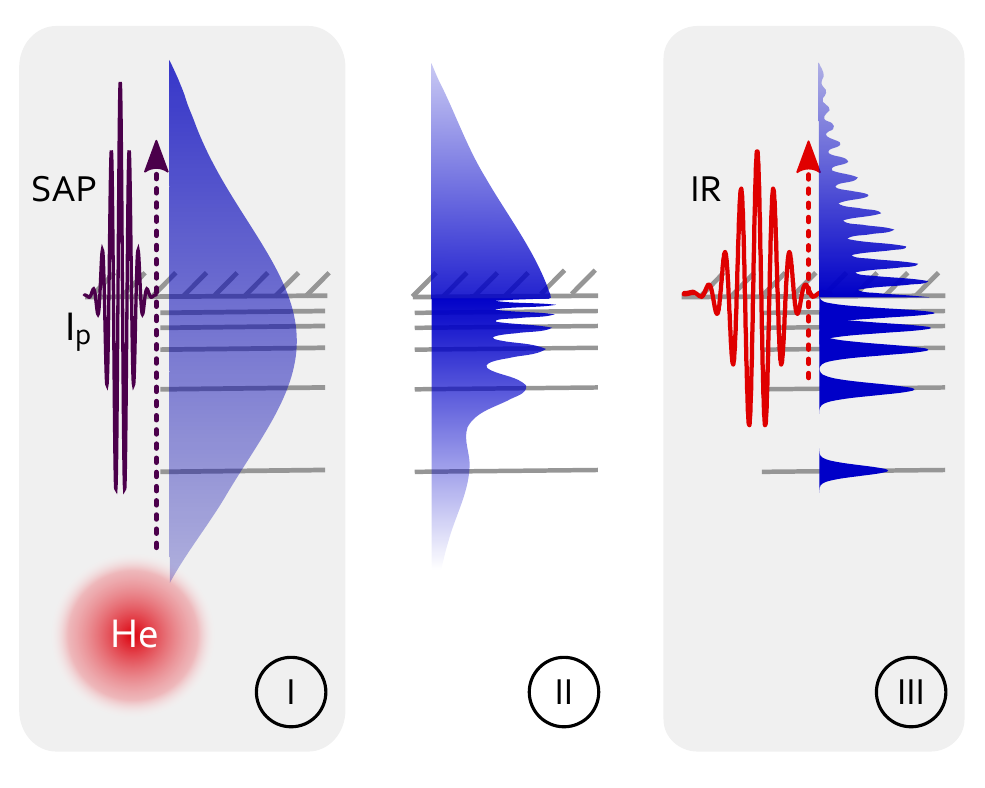}
\caption{(Color online). Principle of attosecond electron interferometry. A broadband, isolated attosecond pulse with a spectrum centered on the ionization threshold of helium is used to coherently excite an electron WP consisting of a superposition of bound and continuum $p$-states (region I). The created WP evolves freely during a certain delay (II). The bound part of the WP is finally ionized by a few-cycle IR pulse (III), which is locked in phase to the attosecond pulse and interferences with the previously created free WP are observed in the photoelectron spectra.}
\end{figure}

Several techniques could be considered to achieve a spectral resolution higher than the inverse of the pulse duration. Although trains of pulses, {\it i.e.} frequency combs have the potential to achieve extremely high spectral resolution (see for example~\cite{SwobodaPRL2010}), pairs of pulses, as in traditional Ramsey spectroscopy, might be easier to implement for time-resolved measurements~\cite{PirriPRA2008}. In this paper we present a novel interferometric technique, using a single attosecond pulse and a delayed IR pulse, that resolves both issues previously discussed: The spectral resolution is much better than the Fourier limit of the exciting pulse and the delayed probe pulse does not perturb the excitation process (see Fig. 1).

In our method, an attosecond pulse is used to excite a bound WP in an atom or a molecule and a delayed IR pulse, locked in phase with the XUV pulse, is used to probe it. Coincident with the creation of the bound WP, we also create a continuum WP, which serves as a reference. After a variable delay, the bound WP is ionized by the IR pulse, and both continuum WPs, created directly by the XUV pulse, or by the two-step (XUV+IR) process, interfere. The analysis of the interferogram obtained when measuring the photoelectron spectrum as a function of delay allows us to determine the spectral components of the bound WP. This technique enables us to obtain a spectral resolution given by the inverse of the IR-XUV delay, which is typically a few tens of femtoseconds \textit{i.e.} more than a factor 100 better than the Fourier limit of the (attosecond) excitation pulse.
We demonstrate the technique experimentally using an attosecond pulse with 350\,as duration and central energy of 24\,eV which excites a broad WP in helium, including bound (unknown) and continuum (reference) components. A few-cycle IR pulse probes the bound WP, and analyzing the electron spectra allows us to recover the composition of the WP.

In the experiments, performed in Milan, a linearly polarized phase stabilized 5-fs IR laser pulse is divided into a central and annular part using a mirror with a hole in the center.
The polarization gating technique is used on the central part to obtain a laser pulse with a temporal window of linear polarization with duration of less than half an optical cycle~\cite{SolaNP2006}. This laser beam is focused into a xenon gas cell to generate XUV radiation via high order harmonic generation. The low order harmonics and the collinear IR radiation are removed using a 100\,nm thick aluminum filter. This metallic filter also provides partial dispersion compensation of the intrinsic positive chirp of the emitted pulse~\cite{LopezMartensPRL2005}, compressing it to a duration of 350 attoseconds. The attosecond pulses are focused using a grazing incidence toroidal mirror into the active region of a velocity map imaging spectrometer (VMIS)~\cite{EppinkRSI1997}, used to record the photoelectron momentum distributions.

The attosecond pulses have a central frequency of 24\,eV with a bandwidth exceeding 10\,eV and excite helium from its ground state to a coherent superposition of bound and continuum $p$~states. At a controllable time-delay the bound WP is ionized by the probe IR laser (bandwidth 0.53\,eV), which is collinearly recombined with the attosecond pulse using a second mirror with a hole in the center and which is focused by a spherical mirror. The IR intensity was close to $1\times10^{13}$ W/cm$^2$, well below that necessary to tunnel ionize He in its ground state, but high enough to induce ``streaking'' when both IR and XUV pulses overlap~\cite{SansoneScience2006,GoulielmakisScience2004}, allowing us to determine the delay between the two pulses. The XUV and IR beams are crossed with an effusive He gas jet emerging from a capillary incorporated into the repeller electrode of the VMIS~\cite{GhafurRSI2009}. Using a set of electrostatic lenses, the electrons emitted in the two-color photoionization process are accelerated onto multichannel plates coupled to a phosphor screen detector. The effusive gas jet allows us to obtain a gas density of $\sim 3\times10^{15}$ cm$^{-3}$ in the interaction region, while keeping a low enough pressure close to the multichannel plates. Two dimensional images are acquired with a CCD camera and used for the retrieval of the 3D initial velocity distribution~\cite{VrakkingRSI2001}.

\begin{figure}[t]\centering
\includegraphics[width=0.8\linewidth]{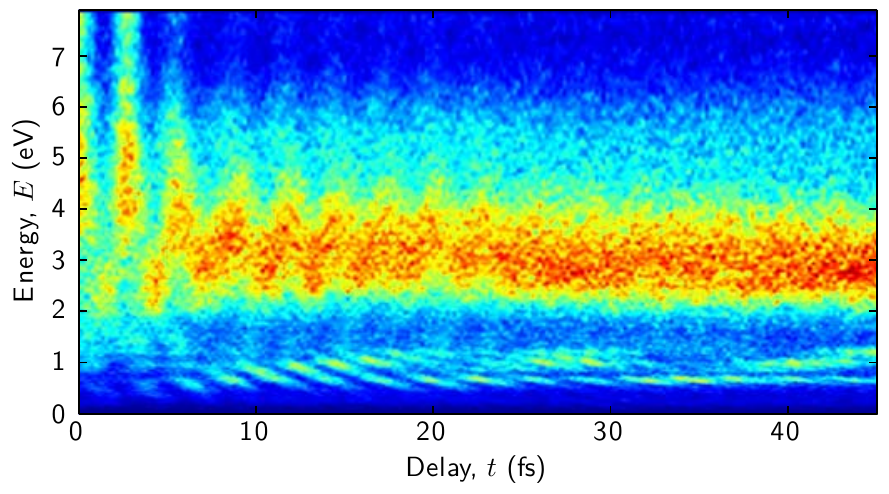}
\caption{(Color online). Experimental photoelectron spectra in He as a function of delay. The electron distribution is recorded using a VMIS and the spectra are obtained by selecting a small collection angle along the polarization direction. The interference fringes are clearly visible at low energies.}
\end{figure}

A scan of the photoelectrons emitted in a small angle around the polarization axis  in the upward direction is shown in Fig.\,2. Both fields are vertically polarized. The color indicates the photoelectron intensity $S_{\mathrm{exp}}(E,t)$ as a function of observation energy $E$ and delay $t$. When the two pulses overlap temporally (at 0\,fs delay), the photoelectron spectra are streaked, indicating photoionization by an isolated attosecond pulse in the presence of a laser field~\cite{SansoneScience2006,GoulielmakisScience2004}. In the more interesting region where the attosecond pulse precedes the IR probe, interference fringes are observed in the low energy region of the spectrum, up to about 2\,eV. As explained below, interferences fringes are expected also at higher energy, but the spectrometer resolution in this region prevents their observation~\cite{note2}. This interference pattern is due to the multiple pathways leading to the same final continuum energy. During the delay, $t$, a continuum state with energy E and a bound, stationary state with energy $E_i$ accumulate a phase difference $(E-E_i )t/\hbar$. The interference fringes, defined as the curves of constant phase difference, are therefore hyperboles, which become more closely spaced as the delay increases.

\begin{figure*}[bth]\centering
\includegraphics[width=0.9\linewidth]{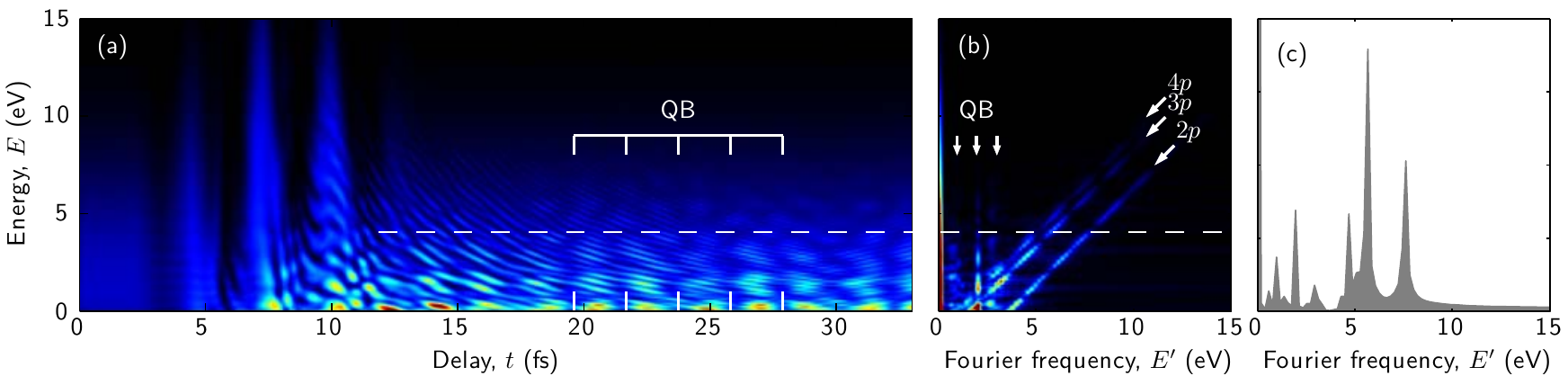}
\caption{(Color online). a) Calculated photoelectron spectra in He as a function of delay between the attosecond pump pulse and the IR probe pulse. Interference fringes are clearly seen at all delays where the attosecond pulse comes before the IR probe. b) Fourier transform of the photoelectron spectrum allowing the identification of the states that form the bound WP. The beat signals from the 2$p$, 3$p$ and 4$p$ states can be seen as vertical lines around 2 eV while the interference between direct and indirect pathways give rise to contributions at an angle of 45 degrees. A lineout of the transform at 4\,eV is presented in c.}
\end{figure*}

In Fig.\,3, we present quantum calculations based on the single active electron approximation~\cite{JohnssonPRL2007} under conditions that are close to those of the experiment. The probe has a 6\,fs duration with a $\cos^2$ shape and an intensity of $1\times10^{13}$ W/cm$^2$, while the duration of the attosecond pulse is 180\,as which is shorter that the 350\,as in the experiment but chosen to match the experimental bandwidth~\cite{note1,note2}. The theoretical spectra as a function of delay, $S_{\mathrm{theo}}(E,t)$, shows electrons emitted in the direction of the IR and XUV polarization axis. Interference fringes appear as soon as the excitation of the bound states is separated in time from their ionization by the probe field, meaning that we have two clearly delineated routes into the continuum. The hyperbolic shape of the fringes is visible in the calculation and the fringe spacing in energy decreases as the delay increases.

A quantum beat signal~\cite{YeazellPRL1988,WoldePRL1988} is also visible in Fig.\,3a, due to simultaneous excitation of several bound states ($i$,$j$). Interferences  between quantum paths leading to the same final energy, for example $E_i+\hbar \omega = E_j+\hbar \omega'$, $\omega$, $\omega'$ being within the IR laser bandwidth, give rise to periodic structures in the ionization probability. The accumulated phase difference between the quantum paths, equal to $(E_j-E_i )t/\hbar$, is independent on the observation energy and the quantum beat signal appears as periodic vertical structures. The beat signal observed in Fig. 3 has a main periodicity of about 2\,fs, corresponding to the beating between the outgoing electrons from the 2$p$ and the 3$p$ states. The periodicity of the beat signal carries information on the relative energy separation of the states involved, while the absolute timing of the beating depends on the relative phase of the pairs of states contributing to the signal. As we now discuss, analysis of the ``direct-indirect" interferences, involving bound states and a reference continuum state allow us to go well beyond quantum beat spectroscopy.

To begin with, the different components of the excited WP can be extracted by Fourier analysis of the delay-dependent photoelectron signal $S_{\mathrm{theo}}(E,t)$. We first analyze the simulation shown in Fig.~3. The Fourier transform at all the possible observation energies yields a two dimensional function of the observation energy $E$ and the Fourier frequency (represented as an energy $E'$). This function, $\cal{S}_{\mathrm{theo}}$$(E,E')$, is presented as a color plot in Fig.~3b  as a function of $E$ and $E'$. Fig. 3(c) shows a lineout at $E=4$ eV. It exhibits six prominent peaks. The three lowest peaks (at approximately 1, 2 and 3 eV) are from the quantum beat between 3$p$-4$p$, 2$p$-3$p$, and 2$p$-4$p$ pairs of states respectively while the other peaks at 4.9, 5.6 and 7.4 eV in Fig. 3c are due to the direct-indirect interferences involving the 4$p$, 3$p$ and 2$p$ states respectively. When the observation energy $E$ is varied, the Fourier frequency of the quantum beats does not change and they appear as vertical lines in Fig. 3b. In contrast, the Fourier frequency $E'$ of the direct-indirect interferences increases with $E$, since the accumulated phase difference between the direct and indirect ionization pathways is proportional to it. This linear relationship results in lines tilted at 45 degrees in Fig. 3b. The energies of the bound intermediate states (4$p$, 3$p$, 2$p$) in the WP can be read directly from the intersections of the 45 degree lines with the horizontal zero energy line. In addition, the relative strengths of the 45 degree lines are directly related to the contributions from each bound state to the ionization signal.

\begin{figure}[t]\centering
\includegraphics[width=0.8\linewidth]{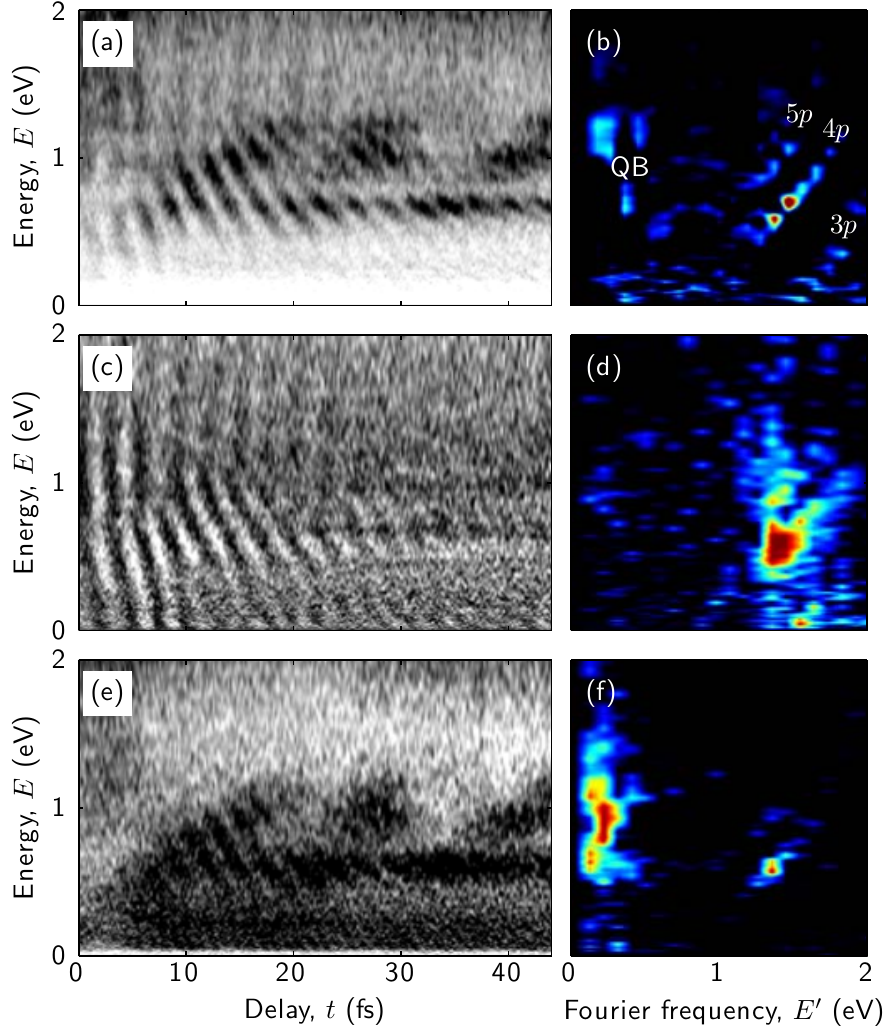}
\caption{(Color online). Analysis of experimental results. (a) Zoom of the low-energy region in Fig. 2. (b) Fourier analysis showing the WP components 3$p$, 4$p$, 5$p$ as well as a quantum beat 4$p$-5$p$. (c) $\beta_1(E,t)$ and (d) its Fourier analysis. (e) $\beta_2(E,t)$ and (f) its Fourier analysis.}
\end{figure}

Turning to the experimental data, we find that quantum beats and direct-indirect interferences are also present. Figure 4 presents an extended analysis of our experimental results. Fig. 4a is a zoom of the low energy region in Fig. 2, while Fig. 4b presents the Fourier transform of the experimental data. The observed 45-degrees lines which are characteristic for the direct-indirect interferences allow us to identify the composition of the bound WP and identify the contributions from the 3$p$, 4$p$ and 5$p$ states. A weak indication of the 4$p$-5$p$ quantum beat can also be seen as a beating with a 13\,fs periodicity in Fig. 4a and as a vertical line near 0.5\,eV in Fig. 4b. Additional information is encoded in the angular distributions since the two interference processes are fundamentally different and involve different angular momentum states. Fig.4 c-f presents a more complete analysis making use of the full angular-resolved photo-electron distribution $F(E,\theta,t)$, which can be expanded as a sum of Legendre polynomials
\begin{equation}
F(E,\theta,t)=\sum_{j=0}^{2L_{max}}\beta_j(E,t) P_J[\cos(\theta)],
\end{equation}
where $P_J$ is the Legendre polynomial of $J^{th}$ order, $\beta_J$ an expansion coefficient, and $L_{max}$ the maximum angular momentum component of the ionized wave function, which corresponds to the maximum absorbed angular momentum. After extraction of the individual expansion coefficients, $\beta_J(E,t)$, we apply the same Fourier analysis that was already applied to the photoelectron spectrum measured along the laser polarization axis. Fig. 4c and 4e shows the extracted expansion coefficients $\beta_1(E,t)$  and $\beta_2(E,t)$ respectively, together with their Fourier transforms in Fig. 4d and 4f. Simple parity arguments dictate that interference between ionization processes that end in states of definite parity (either even or odd with respect to reflection along the polarization direction) will appear in the even $J$ expansion coefficients, while interference between ionization processes that end in mixed parity  states (both even and odd) will appear in the odd $J$ expansion coefficients. The direct-indirect interference process involves electrons that have absorbed different numbers of photons (one for the direct process, two for the indirect process, since the bound states involved can be ionized by absorption of one IR photon) resulting in a mixed parity state.  Indeed, we see that the characteristic hyperbolic interference fringes appear clearly in $\beta_1(E,t)$ (Fig. 4c and d). In contrast, the process that produces quantum beats results in states of definite parity. Accordingly, the quantum beat signal dominates $\beta_2(E,t)$ (Figs. 4e, f), thus confirming the interpretation presented above.

Using the above analysis, we can determine which states are excited by the attosecond pulse (3$p$, 4$p$, 5$p$) and obtain a measure of their relative strengths. Our ultimate goal is of course a complete characterization in both amplitude and phase. Determining the amplitudes and phases of a wave-like object by interference with a reference wave is a well-known approach. The experiment and the analysis we have presented  demonstrates that we can probe an attosecond electron WP using a coherent reference WP that is in the continuum. With this proof of principle in hand, the method can be extended to include the retrieval of a phase associated with each WP component from the interferogram. This phase retrieval, which will be addressed in a future theoretical work~\cite{Mauritsson2010}, yields the phase difference between the two paths into the continuum and includes a phase contribution from both XUV and IR ionization steps.  If it can be arranged that both ionization leading to the creation of the reference WP and ionization by the probe field do not add extra amplitude or phase variation or that these possible variations are well known then a complete phase characterization can be made. Conversely, experiments such as that presented here could be used to characterize the effect of the ionization by the probe field, or some structure in the continuum used for the reference WP.

In conclusion we have demonstrated experimentally a new interferometric technique using a reference continuum WP and a delayed probe excitation. The method demonstrated can be used to probe the temporal evolution of bound or quasi-bound electron WPs, \textit{e.g.} created by shake up~\cite{UiberackerNature2007} with high spectral and temporal resolution simultaneously, thus providing an increased precision when doing attosecond experiments.

\begin{acknowledgments}
This research was supported by the Marie Curie IEF (ATTOCO), RTN (XTRA), EST (MAXLAS) programs, the Integrated Initiative of Infrastructure LASERLAB-EUROPE, the Swedish and European Research Council and the National Science Foundation through grant number PHY-0701372. IZ, SZ, TU and MFK are grateful for support by the DFG through the Emmy-Noether program and the Cluster of Excellence: Munich Center for Advanced Photonics. This work is part of the research program of the "Stichting voor Fundamenteel Onderzoek der Materie" (FOM), which is financially supported by the "Nederlandse Organisatie voor Wetenschappelijk Onderzoek" (NWO). KJS acknowledges support from the Ball Family Professorship.
\end{acknowledgments}


\begin{references}
\bibitem{note2}
The experimental spectrum was also a little modulated, leading to a minimum at 1.5 eV and a maximum at 3 eV.
\bibitem{note1}
In the experiments, the chirp was only partially compensated. In the simulations, we verified that the results did not depend on the chirp but on the bandwidth. We chose to present results obtained with an unchirped pulse.
\end{references}
\end{document}